\documentclass[preprint,preprintnumbers,amssymb,amsmath,9pt, superscriptaddress]{revtex4}[10pt]
\usepackage{graphicx}
\usepackage{dcolumn}
\usepackage{bm}
\begin{document} 
\input epsf.tex
\newcommand{\beq}{\begin{eqnarray}}
\newcommand{\eeq}{\end{eqnarray}}
\newcommand{\nn}{\nonumber}
\def\ltap{\ \raise.3ex\hbox{$<$\kern-.75em\lower1ex\hbox{$\sim$}}\ }
\def\gtap{\ \raise.3ex\hbox{$>$\kern-.75em\lower1ex\hbox{$\sim$}}\ }
\def\CO{{\cal O}}
\def\CL{{\cal L}}
\def\CM{{\cal M}}
\def\tr{{\rm\ Tr}}
\def\CO{{\cal O}}
\def\CL{{\cal L}}
\def\CN{{\cal N}}
\def\mpl{M_{\rm Pl}}
\newcommand{\bel}[1]{\be\label{#1}}
\def\al{\alpha}
\def\bt{\beta}
\def\eps{\epsilon}
\def\eg{{\it e.g.}}
\def\ie{{\it i.e.}}
\def\mn{{\mu\nu}}
\newcommand{\rep}[1]{{\bf #1}}
\def\be{\begin{equation}}
\def\ee{\end{equation}}
\def\bea{\begin{eqnarray}}
\def\eea{\end{eqnarray}}
\newcommand{\eref}[1]{(\ref{#1})}
\newcommand{\Eref}[1]{Eq.~(\ref{#1})}
\newcommand{\gsim}{ \mathop{}_{\textstyle \sim}^{\textstyle >} }
\newcommand{\lsim}{ \mathop{}_{\textstyle \sim}^{\textstyle <} }
\newcommand{\vev}[1]{ \left\langle {#1} \right\rangle }
\newcommand{\bra}[1]{ \langle {#1} | }
\newcommand{\ket}[1]{ | {#1} \rangle }
\newcommand{\eV}{{\rm eV}}
\newcommand{\ev}{{\rm eV}}
\newcommand{\kev}{{\rm keV}}
\newcommand{\Mev}{{\rm MeV}}
\newcommand{\gev}{{\rm GeV}}
\newcommand{\tev}{{\rm TeV}}
\newcommand{\mev}{{\rm MeV}}
\newcommand{\meV}{{\rm meV}}
\newcommand{\mnu}{\ensuremath{m_\nu}}
\newcommand{\nnu}{\ensuremath{n_\nu}}
\newcommand{\mlr}{\ensuremath{m_{lr}}}
\newcommand{\acc}{\ensuremath{{\cal A}}}
\newcommand{\mav}{MaVaNs}
\title{Supersymmetric Theories of Neutrino Dark Energy}
\author{Rob Fardon}
\affiliation{
Department of Physics, Box 1560, University of Washington,\\
           Seattle, WA 98195-1560, USA
}\author{Ann E. Nelson}
\affiliation{
Department of Physics, Box 1560, University of Washington,\\
           Seattle, WA 98195-1560, USA
}\author{Neal Weiner}
\affiliation{Center for Cosmology and Particle Physics,\\
  Department of Physics, New York University,\\
New York, NY 10003, USA}
\preprint{ }
\begin{abstract}
We present a supersymmetric model of dark energy from Mass Varying Neutrinos which is stable against radiative corrections to masses and couplings, and free of dynamical instabilities. This is the only such model of dark energy involving fields with significant couplings to any standard model particle. We briefly discuss consequences for neutrino oscillations and solar neutrinos.
\end{abstract}
\maketitle
\section{Introduction}
The precision cosmological data acquired over the last decade has changed the nature of theoretical cosmology. Models which were previously viable, such as hot or warm dark matter, or the open universe model, have been excluded by experiments such as SDSS\cite{Abazajian:2004aj}, WMAP\cite{Spergel:2003cb}, 2dF\cite{Percival:2001hw}, and many others \cite{deBernardis:2000gy,Lee:2001yp,Halverson:2001yy}. One of the most exciting aspects of the new precision cosmology is that we are using gravity to  learn  a great deal about the dominant constituents of the universe. The tremendous results of the High-z Supernovae Search Team \cite{Riess:1998cb}, as well as the Supernova Cosmology Project (SCP) \cite{Perlmutter:1998np}, have given solid evidence that the universe is accelerating, potentially due to a new, negative pressure fluid.

This new fluid, unlike dark matter, does not possess an obvious candidate  for any observable  interactions with standard model particles. In fact, the most popular explanation - a cosmological constant - has no dynamics associated with it at all. Slow-roll quintessence offers the possibility of detecting an equation of state different from $w=-1$, but little else in the way of detectable signals. Furthermore, the slow-roll condition requires a field with mass on the order of the Hubble scale, $10^{-33}\ \ev$. If the quintessence field couples to   to standard model fields,  with couplings of gravitational strength, the typical size of quantum radiative corrections to the scalar potential are at least $10^{60}$ times too large, unless exquisitely precise  cancellations occur. 

Are there theories of dark energy which have naturally sized quantum corrections yet have possible non-gravitational signatures (in particular, those arising from couplings to standard model fields)?
We are willing to  assume here the existence of some unknown, perhaps nonlocal, physical mechanism which can ensure zero vacuum energy, despite the apparently enormous quantum contributions. We still may ask whether, given this assumption, a model of dark energy in which the radiative corrections to mass and coupling parameters of the  model do not have to be finely tuned against the tree level terms.

In \cite{us}, it was suggested that relic neutrinos could form a negative pressure fluid, and yield  cosmic acceleration. Such a neutrino fluid could occur if neutrinos interact through a new scalar force. The idea is appealing because the neutrino mass   scale is comparable to that of the dark energy, and the relic neutrinos form a smooth background which we expect to permeate the present universe. The result of the scalar force is that the neutrino mass depends on neutrino number density, and thus evolves on cosmological timescales.

New scalar forces for neutrinos have been considered for quite some time, in the context of neutrino oscillations \cite{wolfenstein, Sawyer:1998ac}, neutrino dark matter \cite{Kawasaki:1991gn}, and neutrino clouds \cite{Stephenson:1996qj}.

With dark energy arising from such Mass Varying Neutrinos (MaVaNs), a O($10^{-33}\ \ev$) mass scalar field is not required. The mass of the scalar field can be as large as  O($10^{-4}\ \ev$) - the scale of the neutrino number density - in order to employ a mean-field approach. This mass scale is both  comparable to the dark energy scale, and more plausibly stable against radiative corrections than the Hubble scale. In ref. \cite{us2} it was argued that   the local neutrino mass could also depend on the local   matter density in this scenario, offering the possibility that neutrino oscillation experiments could shed light on the nature of dark energy. Much subsequent work has been done on the phenomenology of MaVaNs, including effects on leptogenesis \cite{Bi:2003yr}, high energy astrophysical neutrinos \cite{Hung:2003jb}, gamma ray bursts \cite{Li:2004tq}, solar neutrinos \cite{Barger:2005mn,Cirelli:2005sg}, the cosmology of particular forms of the potential \cite{Bi:2004ns,Brookfield:2005td}, and studies of models of various types \cite{Peccei:2004sz}. Note that other models have attempted to relate dark energy and neutrino mass in slow-roll quintessence theories \cite{Hill:1988bu,Hung:2000yg,Barbieri:2005gj}, modifications of the energy momentum tensor \cite{Guendelman:2004hk} and a metastable vacuum energy \cite{Singh:1994nt}. However, these do not rely upon a new, milli-eV scale force between the relic neutrinos, and cannot provide the  interesting phenomenology of refs~\cite{us,us2}, namely, that neutrino mass depends on the environment.

While MaVaN dark energy is an appealing framework, ref. \cite{us}, does not present a  complete, radiatively stable model, and questions about the scalar potential and its origins remain. In this paper, we provide such a radiatively stable theory of MaVaN dark energy, which admits additional interactions with standard model fermions,  explains the origin of the size of the dark energy in terms of   neutrino mass parameters, and  can easily mesh with a comprehensive effective description of physics below the Planck scale. In section \ref{sec:review}, we review the basic features of MaVaN dark energy, including the associated theoretical problems. In section \ref{sec:susyhybrid}, we will show that including supersymmetry can not only address the quantum corrections, but naturally leads one to consider a new class of dark energy models which are analogous to the ``hybrid'' inflation models \cite{linde}. In section \ref{sec:threenu}, we describe a complete model of neutrino physics, with three additional sterile neutrinos, and discuss the consequences for upcoming experiments. In section \ref{sec:pheno}, we consider the phenomenological implications of this scenario, in particular for solar neutrinos. Finally, in section \ref{sec:conclusions}, we review the scenario, and discuss future directions.

\section{Dark Energy from Mass Varying Neutrinos}
\label{sec:review}

In \cite{us} it was proposed that neutrinos interacting via a new scalar force could act as a form of dark energy. The idea is most simply understood by promoting the neutrino mass, \mnu, to a dynamical field, which, itself, has an associated scalar potential. The energy of the system is then
\be
V=\mnu \nnu + V_0(\mnu).
\ee
Note that \mnu\ need not be a canonically normalized field.
Assuming the curvature scale of the potential is much larger than the Hubble expansion rate, we can use an adiabatic solution to the equations of motion where   \mnu\  tracks the point which instantaneously minimizes the total energy, that is, where
\be
V'=\nnu + V_0'(\mnu) = 0.
\ee
One can then show
\be
w+1 = -\frac{V'(\mnu)}{V{\mnu}} =\frac{ \Omega_\nu}{\Omega_\nu+\Omega_{DE}},
\ee
where $\Omega_{DE}$ is the energy density stored in the scalar potential, and $\Omega_\nu$ is the neutrino energy density.

This can be simply illustrated in a concrete model. We begin by including a scalar field, \acc, which is the canonically normalized field responsible for the dynamical neutrino mass, and a fermion $N$. 
Because the stored potential energy of \acc\ is responsible for the acceleration of the universe, we called this field the ``acceleron''. $N$ is taken to be  a left chiral Weyl field with no Standard Model gauge charges, which may be referred to as a ``dark'', ``right-handed'', or ``sterile'' neutrino.

We then consider the interactions,
\be
{\cal L} \supset m_D \nu N + \kappa \acc NN + {\rm h.c.} + V_0(\acc) \ ,
\label{eq:nonsusylagrangian}
\ee
where  $\kappa$ is some Yukawa coupling, and $\nu$  is a two-component left-chiral Weyl field describing some linear combination of the active neutrinos. This Lagrangian is appropriate for energy scales well below 100 GeV, as we have integrated out the Higgs field.
If $\kappa \acc \gg m_D$, we can remove $N$ from the low energy effective theory and are left with 
\be
{\cal L} \supset \frac{m_D^2}{\kappa \acc} \nu_l^2 + {\rm h.c.}+ V_0(\acc).
\ee
If the relic neutrinos are light mass eigenstates, then we must consider the system at finite density, where we have an effective potential for \acc,
\be
V_{\rm eff} = n_\nu \left|\frac{m_D^2}{\kappa A}\right|+ V_0(\acc).
\ee
Because the first term tends to drive \acc\ to larger values, the result of this effective potential is to have a neutrino number density dependent value for the \acc\ condensate, even if the vacuum expectation value is zero.

A very flat scalar potential $V_0$, such as a logarithm or a small fractional power, will give an equation of state parameter for the acceleron-neutrino fluid which is close to -1, as is phenomenologically required for dark energy.

The framework suffered from two principle theoretical shortcomings. First, a flat potential was needed, and both quadratically- and logarithmically divergent radiative corrections were too large unless new states appeared in the theory at a scale of order $10^{-2}\ \ev$. Second, a quadratic potential was not flat enough to give a dark energy with equation of state $w \simeq -1$, and the necessary small fractional power law, or logarithmic, potential  had no obvious microscopic origin. It has also been recently argued that such theories suffer from instabilities at late times \cite{matias}. We will return to this point later.  

 There are a number of known possibilities to control the size of radiative corrections such as supersymmetry, compositeness, extra dimensions, strong near-conformal interactions,  and pseudo-Goldstone boson type shift symmetries. In this paper, we will focus on supersymmetry, which, as a perturbative theory, allows us to consider the theory at arbitrarily high temperatures in a theoretically controlled way. 

The question of the origin of the shape of the potential is more difficult. Logarithmic potentials do arise in gauge mediated supersymmetry breaking \cite{Dine:1993yw,Dine:1994vc,Dine:1995ag,Giudice:1998bp,deGouvea:1997tn}, but only at vevs larger than the supersymmetry breaking scale. However, we shall see that a mildly different approach will free us of the need for such flat potentials.

\subsection{Hybrid models}
Many of the theoretical questions associated with dark energy are similar to questions that have arisen in the context of inflation. 
One highly successful framework has been that of hybrid inflation \cite{linde}. In hybrid inflation, the energy density which drives inflation resides in the potential for some field which is   in a false minimum due to the large value of another  light field.  A typical hybrid model has a potential
\be\label{hybridpotential}
V = m^2 \acc^2 + \kappa^2(\CN^2 - v^2)^2 + \alpha  \acc^2 \CN^2 \ . 
\ee
In hybrid inflation, \acc\ is referred to as the slow-roll field, while $\CN$ is referred to  as the waterfall field. We use \acc\   to make a connection  with MaVaN theories.  As long as \acc\  has a large expectation value,  $\CN$  has a positive mass squared. When the value of \acc\ becomes smaller than a critical value $\acc_c = 2 \kappa^2 v^2/\alpha $,   $\CN$ becomes tachyonic and will begin to roll.

In MaVaN theories, rather than being a slowly rolling field, \acc\ will have a large expectation value due to the presence of relic neutrinos. However, in analogy with the hybrid inflation theory, the tachyonic field $\CN$ will be stabilized in a false minimum at $\CN=0$, leaving a vacuum energy $\kappa^2 v^4$. So long as the energy density stored in the \acc\ condensate is sufficiently small, the combined scalar potential will appear as a dark energy with equation of state $w\approx-1$.

\section{A Supersymmetric Hybrid Model}
\label{sec:susyhybrid}
Let us now combine the disparate elements of MaVaNs, supersymmetry, and hybrid inflation. Remarkably, in a minimal supersymmetrization of  the simplest MaVaN model, we   immediately arrive at a hybrid model, with the only assumption being the sign of the supersymmetry breaking mass squared term for the sterile sneutrino.

We begin with the simplest Lagrangian for a MaVaN theory, stated in eq. \ref{eq:nonsusylagrangian}.
We promote $\nu$, $N$ and \acc\ into chiral superfields   $\ell$, $n$ and $a$. The fermionic interactions of eq. \ref{eq:nonsusylagrangian} are easily captured in the following superpotential 
\be
W = \kappa a n n  + m_D \ell n \  .
\ee
The resulting scalar potential   is
\be
V= 4 \kappa^2 \acc^2 \CN^2 +\kappa^2 \CN^4 + m_D^2 \CN^2 \ ,
\ee
where $\CN$ is the scalar component of $n$. Here we consider only the light fields in the theory, in particular, we set ${\cal L}$ to zero as the sneutrino has a large soft mass.
Up to a constant, this potential bears a remarkable resemblance to eq. \ref{hybridpotential}, other than the sign of the $\CN$ mass term.  However, inclusion of supersymmetry breaking corrections can easily change this sign.  

Let us consider the supersymmetry breaking quantum corrections to this potential. To begin, we should consider radiative corrections above the scale of electroweak symmetry breaking. We expect the Dirac mass term to arise from a small Yukawa coupling, $W\supset y_D H LN$. The one loop   radiative correction to the $\CN$ mass is 
\be
\delta(\mu> v) m_n^2 \sim -\frac{8 y_D^2 m_{\rm susy}^2}{16 \pi^2} \log(\Lambda/v)\ ,
\ee
where $\Lambda$ is the  scale at which the soft masses get generated, and $m_{\rm susy}^2$ is the average of the soft susy breaking higgs and slepton masses squared. 
Note that  for   typical values of $\Lambda$ and $m_{\rm susy}^2$, the natural size of this correction is  of the order $m_D^2$, and the sign is negative.

The susy breaking contributions to the $\CN$  mass squared below the electroweak scale are all proportional to $\kappa^2 m_D^2$, which is in general smaller than $m_D^2$, even when log enhanced. 

The acceleron field receives a log enhanced   correction to its mass squared of
\be
\delta\tilde{m}_\acc^2 \approx \frac{\kappa^2 }{16 \pi^2}\left(  2  m_D^2\log (\tilde m_{\tilde \nu}^2/m_D^2)- \delta \tilde{m}_{\CN}^2\log{(\Lambda^2/m_D^2)}\right)\,
\ee 
where $\tilde{m}_{\tilde \nu}$ is the mass of the active sneutrino, $\delta \tilde m_{\CN}$ is the susy breaking mass of the sterile sneutrino, and $\Lambda$ is the scale at which $\delta \tilde m_{\CN}$ is generated.
With reasonable values for $m_D$, $\Lambda$, and $m_{\tilde \nu}^2$, this is roughly $m_\acc^2 \simeq 0.4  \kappa^2 (2 m_D^2 - \delta m_n^2)$. 

Of course, there are also gravity mediated and Planck scale mediated effects whose size depend on the gravitino mass and on the degree of sequestering \cite{Randall:1998uk} of the $a,n$ sector from the supersymmetry breaking sector.
 We take the  calculations of radiative corrections as giving a rough lower bound on the natural size for the magnitudes of the soft susy breaking masses squared of \acc\ and $\CN$.  

Altogether, the potential at low energies is
\be
V = -c_1  m_D^2 \CN \CN^* +\kappa^2  
\CN^2 \CN^{*2} + 4 \kappa^2 \acc \acc^* \CN \CN^* +c_2\kappa^2 m_D^2 \acc\acc^* + {\rm constant},
\ee
where we have assumed the UV physics achieves the appropriate signs, and the $c_i$ are coefficients of order unity. (Later, we shall see that $\kappa$ is a small parameter, hence we ignore terms $O(\kappa^4)$, such as one-loop radiative corrections to the quartic couplings.)

We can now consider the late time dynamics of this theory, with naturally sized parameters.
We choose the constant in  the potential to set the true vacuum energy density to zero.
The true minimum of this potential is at $\vev{\acc}=0$ and$|\vev{\CN}|=c_1^{1/2}m_D/\sqrt{2} \kappa$, and the false minimum, $\CN=0$, has energy density $c_1^2 m_D^4/4 \kappa^2$. Since there are no gauge interactions to maintain a large $\kappa$ in the IR, it is natural to assume that $\kappa$ is small. If we wish to identify this potential contribution as the present dark energy density, we can thus bound $m_D$ by naturalness to be
\be
m_D \lsim 10^{-2.5}\ \ev.
\ee
Thus, naturalness allows us to identify this as the Dirac mass explaining the solar splitting, or some lighter mass, but would involve a one part in $10^{3}$ to $10^4$ tuning if we wished to identify $m_D$ as the mass associated with the atmospheric splitting.

If we assume that the relic neutrinos are in the light mass eigenstate, then we can estimate the size of the \acc\ vev,
\be
V_{\rm eff} \simeq {3\zeta(3)T^3  m_D^2 \over 2 \pi^2 \kappa |\acc|} + c_2 \kappa^2 m_D^2 \acc \acc^*
\ee
in the non-relativistic case and
\be
V_{\rm eff} \simeq {T^2  m_D^4 \over 12 \kappa^2 |\acc|^2} + c_2 \kappa^2 m_D^2 \acc \acc^*
\ee
in the relativistic case.
Minimizing the effective potential yields
\be
\kappa \acc \simeq T/2 c_2^{1/3}
\ee
in the non-relativistic case, and
\be
\kappa \acc \simeq \sqrt{m_D T/12 c_2^{1/2}}
\ee
in the relativistic case. 
We require $\kappa \acc \gsim m_D$, both so the seesaw Majorana mass formula works, and, more importantly, so that $\CN$ has a positive mass-squared. This, in turn, requires either $c_2 \ll 1$ (an unnatural tuning) or $m_D  \lsim T$ (meaning the neutrinos are at least moderately relativistic). Consequently, naturalness will force the neutrino responsible for the dark energy to be the lightest of the three neutrinos, and relativistic. While we might enjoy making a direct connection with known neutrino masses, both the scale of dark energy and the need for a metastable minimum point  us towards the lightest neutrino as the origin of dark energy.  Note that such theories do not suffer from the potential instabilities of a highly non-relativistic neutrino dark energy theory \cite{matias}.

\subsection{Multiple neutrinos}
\label{sec:threenu}
Since there are   three active neutrinos, and evidence for  at least two mass scales above $10^{-4}\ev^2$,  we would like to extend this theory to include  all three neutrinos and an explanation for the atmospheric and solar neutrino oscillations. We restrict ourselves to a single acceleron, and three dark fermions $N_i$. In doing so we should extend the superpotential to
\bea
W = \kappa_{ij} a n_i n_j + m_{i} n_i \ell_i.
\eea
Here we have written the superpotential in the basis where the Dirac mass matrix is diagonal, with eigenvalues ordered $m_3>m_2>m_1$. The renormalization of the \acc\ mass takes place as in the previous section, but with additional diagrams possible from the off-diagonal couplings between \acc\ and the right handed neutrinos. 

The same considerations of naturalness apply here, and we require that the lightest mass eigenstate be responsible for maintaining the present false minimum for the \acc\ field, and, more specifically, that
\be
\kappa_{ij}^2 m_i^2 > \kappa_{11}^2 m_{1}^2
\ee
where $i\ge j$ and $i>1$.

Making this somewhat more quantitative, since $T_\nu \simeq 10^{-4}\ev$, we must have $m_1 \lsim 10^{-4} \ev$. To achieve the proper vacuum energy, this requires $\kappa_{11} \lsim 10^{-3}$. As a consequence, we expect an acceleron mass $m_\acc \sim 10^{-7} \ev$. Taking $m_2 \ge 10^{-2}\ev$ as necessary to explain the solar neutrino deficit, and $m_3 \ge 10^{1.5}\ev$ as necessary to explain the atmospheric neutrino deficit, we have $\kappa_{2 a} \lsim 10^{-5}$ and $\kappa_{3a} \lsim 10^{-5.5}$.

If the radiative corrections to all the $\CN$ soft masses are simply proportional to $y_D^2$, then all masses squared must be negative. If this is the case, then $n_3$ and $n_2$ would have rolled off by now, achieving vevs $n_i \sim m_i/\kappa_{ii}$ (assuming it is stabilized by the quartic piece), resulting in an acceleron mass $O(m_i)$. Since the acceleron Compton wavelength must be longer than the cosmic neutrino  separation, such a large vev is unacceptable. To reduce the $\CN$ expectation value, 
one can include terms 
\be
\lambda_{ijk} n_{i}n_jn_k.
\ee
These need only be large for $n_{2,3}$ in order to prevent a large \acc\ mass, and for our purposes, we will assume that couplings $\lambda_{1ij}$ to the lightest generation are small.

Alternatively, it is possible that the corrections to the soft masses squared are not simply proportional to $y_D^2$. One could then have    positive masses squared for the heavier two sterile sneutrinos, but a negative mass squared for $\CN_1$. This could arise, for instance, if the heavier active sneutrinos are non-degenerate, or in the presence of  small corrections form Planck scale physics.

\subsection{Origin of the Small Couplings}
Our model requires the introduction of several small dimensionless numbers, such as the Yukawa couplings which give neutrinos Dirac masses of less than an eV, and the parameters $\kappa_{ij}$. In addition, we have left out several couplings which are not necessarily forbidden by symmetries, such as a superpotential term $a h_u h_d$, but whose coefficients must be sufficiently   small.
A large number of explanations for such small or absent parameters exist in the literature. For instance, one could have a new dimension of size somewhat larger than the Planck length, with standard model fields localized to a brane on one end, the acceleron on a different brane, and the 
sterile neutrinos in the bulk, explaining why the acceleron has the most suppressed direct couplings to matter and why the couplings of the sterile neutrinos have various suppression factors \cite{Randall:1998uk,Arkani-Hamed:1999dc}. In the version of the model with no sterile sneutrino vevs, a continuous or discrete lepton number symmetry could suppress   unwanted couplings such as $a h_u h_d$ and  
$n h_u h_d$. 
\subsection{The Universe from $T\sim 1\ \ev$ to the Present}
We will assume that the universe is populated by mass eigenstates after BBN. The circumstances under which this occurs we will leave for discussion elsewhere \cite{Weiner:2005ac}. 

If the masses-squared for the heavier scalars are positive, then the fluid behaves as a traditional hot MaVaN background, but with a quadratic potential, an energy density which redshifts far more rapidly than dark energy.

If the masses-squared for the heavier scalars are negative, then at some critical temperature, the $\CN_{2,3}$ fields will become tachyonic and roll from the false minimum. 

The exact vevs and the overall phenomenology is a complicated function of $\lambda_{ijk}$. However, we will in general require $\lambda_{1jk} \ll \lambda_{ijk \ne 1}$ so that the lightest sneutrino does not acquire a large mass when the more massive sneutrinos acquire vevs. For instance, let us consider the three-neutrino terms
\be
W\supset \lambda (n_2^3+n_2^2 n_3 + n_2 n_3^2+n_3^3)
\ee
We take the form of the couplings purely for illustration. If we take $\tilde m_{3}^2 \gg \tilde m_2^2$, then we can first focus on the early universe behavior of the heaviest sneutrino.

The scalar potential (neglecting couplings to the acceleron) is
\be
V= -\tilde m_3^2 |\CN_3^2| + \lambda^2 |3\CN_2^2+ 2\CN_2 \CN_3+ \CN_3^2|^2 + |3\CN_3^2+ 2\CN_2 \CN_3+ \CN_2^2|\ .
\ee
One can show this is minimized for $\CN_2 = \mp 0.19 \tilde m_3/\lambda $, $\CN_3 =\pm .3\tilde m_3/\lambda$.
The resulting neutrino mass matrix for the heavier two neutrinos is
\be\label{samplemass}
M_\nu = \begin{pmatrix} 0 & 0 & m_2 & 0 \\ 0 & 0 & 0 & m_3 \\ m_2 & 0 & 0.6\ \tilde m_3 & 0.6\ \tilde m_3 \\ 0 & m_3 & 0.6\ \tilde m_3 & 1.8\ \tilde m_3
\end{pmatrix}
\ee
Under the assumption that $\tilde m_3 \sim m_3$,  there are two mostly sterile Majorana neutrinos, with masses of   1.8 $\tilde m_3$ and 0.6 $\tilde m_3$.  A neutrino with   mass  of order $0.6  m_3^2/\tilde{m}_3$ is mostly active, but with a moderate sterile component. We associate this mass scale with atmospheric neutrino oscillations, so the active component is almost entirely $\mu $ and $\tau$. There is a  $ 1.7 m_2^2/\tilde{m}_3$ mass neutrino which has very small sterile component, which we associate with solar neutrino oscillations.  
The vacuum energy of the metastable state is $\Lambda^4 = 0.046 \tilde m_3^4/\lambda^2$. If we take $\tilde m_3 \sim m_3 \sim .05 \ev$, the energy density is $O((2 \times 10^{-2})^4/\lambda^2)$.

We now must determine the temperature at which the heavier sneutrino vevs roll to their minimum. We will consider the period when all neutrinos are relativistic.
Taking the acceleron Yukawa matrix to be diagonal, the vev of the acceleron will be determined by its mass (which we take by naturalness to be $O(\kappa_1 m_1)$, and the larger of $m_i^4/\kappa_i^2$ for $i=2,3$. The effective potential is
\be
V\sim T^2 \Sigma_i \frac{m_i^4}{\kappa_i^2 \acc^2} + m_\acc^2 \acc^2.
\ee
If $m_3^4/\kappa_3^2 > m_2^4/\kappa_2$, then $(\kappa_3 \acc)^4 \sim \frac{\kappa_3^4 T^2 m_3^4}{\kappa_1^2 m_1^2} \lsim T^2 m_3^2$, in which case, the heaviest sneutrino would become tachyonic no later than $T \simeq m_3$. On the other hand, if $m_2^4/\kappa_2^2> m_3^4/\kappa_3^2$, then this can last until $T\sim m_2$, when the intermediate mass neutrinos become non-relativistic. In either event, the heavier sterile sneutrino vevs should roll from the metastable minimum no later than roughly $T\sim 10^{-2}\ \ev$ (using the solar neutrino data as a guide), but possibly much earlier than $T \sim .05\ \ev$. 

As a consequence, there will be energy in the   oscillating heavier sterile sneutrino field. Such scalar field dark matter   will cluster like cold dark matter, but with fluctuations smoothed out on the  neutrino free-streaming length at the time when the field begins to oscillate. It seems difficult to come up with a viable scenario where the entire dark matter component of the universe is dominated by such oscillations. 
It is   easy to find parameters where the energy density of the oscillating scalar field is much smaller than that of dark matter. 
Even a subdominant dark matter component is very interesting, as it could potentially be observed or constrained from studies of the matter power spectrum.  Such an analysis is beyond the scope of this paper. 

Besides the four neutrinos already discussed, the resulting  neutrino spectrum will have   one Majorana MaVaN pair, whose mass continues to evolve on cosmological times. The lightest sneutrino remains in its metastable minimum.

\section{Phenomenology}
\label{sec:pheno}
The neutrino mass phenomenology of this scenario is potentially very rich. With the additional singlet states, one must be concerned with active--sterile oscillations, as well as active--active, and with the light acceleron and sterile sneutrino fields, novel matter effects are possible. We will discuss the two distinct scenarios outlined above: namely, case 1, in which there are no sneutrino vevs, has a set of two, heavier, pseudo-Dirac neutrinos with a light pair of weakly-Majorana neutrinos, and case 2, which has sneutrino vevs,    a heavy and light pair of weakly-Majorana neutrinos, and an intermediate mass Majorana neutrino whose mass comes from a singlet with a mass of the atmospheric scale. 

In the case 1), we shall see that the splittings of the heavier pseudo-Dirac neutrinos can be made sufficiently small that there are no significant consequences, but the splitting of the lightest, weakly Majorana  neutrinos allows for an MSW enhanced   conversion of low energy solar neutrinos into sterile states, which must be avoided. We will see that for sufficiently small splitting, this need not be a concern, or, in the presence of acceleron-matter couplings, the singlet state can be made sufficiently heavy in the sun to prevent level crossings.

In the case 2), we shall see that the high energy solar neutrinos are not significantly changed so long as the singlet neutrinos acquiring masses from sneutrino vevs have masses at least $0.03\ \ev$, so as to prevent level crossings with high energy neutrinos. All mixing angles involving the two heavier sterile neutrinos are fairly small, but it is possible for the  new sterile states  to affect  long-baseline neutrino oscillation experiments. In particular, there could be a moderate sterile component in the oscillations of   muon neutrinos, with a somewhat shorter oscillation length than that of $\mu-\tau$ oscillations. 

There have been earlier studies of MaVaNs in the sun. In particular, \cite{Cirelli:2005sg} studied the MaVaN dark energy scenario with the sterile states heavier than an MeV, while \cite{Barger:2005mn} studied the effects of a particular model in the sun, finding better agreement with data than the standard LMA result. However, neither analysis applies to the present discussion.

There are many other scenarios to consider which we do not discuss further. For instance, one could allow direct coupling of the sterile sneutrino fields to ordinary matter. In this case the sterile sneutrinos could acquire expectation values which depend on the matter density. Since the sterile sneutrino vevs appear in neutrino mass matrices, this  could  give novel matter effects in neutrino oscillations, as in \cite{us2} and discussed in \cite{Zurek:2004vd}, such as a potential explanation of the LSND anomaly \cite{Athanassopoulos:1998pv}.

\subsection{Solar Neutrinos}
\label{sec:solar}
In the previous section we saw that a natural theory of dark energy involved three active and at least three sterile neutrinos.  A natural dark energy mechanism  involves  one active and one sterile neutrino, with Majorana  masses in the $10^{-4}$ eV range. The observed atmospheric and solar neutrino oscillations  are produced by larger Dirac mass terms. 
In case 1, 
none of the oscillations into  sterile neutrinos affect   terrestrial neutrino experiments, as the associated oscillation lengths are between $10^5$ and $10^8$  kilometers for typical   neutrino energies. However, terrestrial  detection of solar neutrinos could be affected by oscillations  into sterile neutrinos. 
For solar neutrino phenomena  we focus on case 1, as this is potentially more dangerous. In this section we discuss solar neutrino phenomena for a   particular spectra. We  argue that the electron neutrinos produced by the $B^8$ process observed by the SNO and superK experiments are mostly converted to a linear combination of $\mu$ and $\tau$ neutrinos, while lower energy solar electron neutrinos are partially  converted to sterile neutrinos.

 The terms in an effective Lagrangian involving neutrinos and the acceleron are

\be
\CL\supset m_{ai}  N_a \nu_i+   \kappa_{ab} N_a N_b \acc +{\rm h.c.}+ \frac{m^2_\acc}{2} \acc^2
\ee
Here  $a=1,2,3$ is an index labelling the sterile neutrinos, and  $i=e,\mu,\tau$ labels the active neutrinos.

The Dirac type mass matrix $m$ is taken to have eigenvalues 
$m_3\approx 4\times 10^{-2}$~eV, 
$m_2\approx 9\times 10^{-3}$~eV, and 
$m_1\sim 10^{-4}$~eV.   A natural dark energy model may be found where the Majorana-type matrix 
$k_{ab}\vev\acc$
has no eigenvalues larger than of order 
$10^{-4}$~ eV,  has an acceleron vev 
$\vev{\acc}\sim 10^{-1.0}$~eV, an acceleron mass 
$m_\acc\sim 10^{-7}$~eV, and in a basis where $m$ is diagonal, 
$k_{11}\sim 10^{-3}$,   $k_{3,a}\lesssim 10^{-5.5}$ and 
$k_{2,a}\lesssim 10^{-5}$.   With three sterile neutrinos, in case 1 there will be two pairs of pseudo-Dirac neutrinos with masses of order 
$m_{2,3}$ and mass splittings which are less than of order 
$10^{-7}$~eV, and a pair of MaVaN Majorana neutrinos with masses and mass splittings of  order $10^{-4}$~eV.  The mass $m_2$ is associated with solar neutrino oscillations   and  $m_3$ with atmospheric. Because the SNO  and SuperK  data \cite{sno,super} indicate that the 
more energetic  solar neutrinos do not undergo large oscillations into sterile  states on their way to the earth, we will assume that the mass squared splitting of the second pair is   less than of order 
$10^{-12}$~eV${}^2$, so that the associated oscillation length  is  greater than 1AU for solar neutrinos. We therefore require 
$k_{22} \lesssim  10^{-8}$.

Due to the MSW effect \cite{wolfenstein,Mikheev:1986gs}, the high energy neutrinos,  with energies above 6 MeV, are then mostly adiabatically converted into the active component of the second Dirac pair as they propagate from the center of the sun.  This component is not exactly a mass eigenstate in vacuum, but the   splitting between the mass eigenstates is  too small to affect neutrino oscillations at distances less than many AU. 

The low energy solar neutrinos are another story.  The SAGE \cite{Abdurashitov:1999zd} and GALLEX \cite{Hampel:1998xg,Altmann:2000ft} experiments have detected  solar  neutrinos with energies from  0.2 MeV, and, when combined with other solar neutrino experiments, the results indicate that the survival probability of electron neutrinos in the 0.3  MeV energy range should be above 0.47 to be within the  2 sigma error bars of the experimental data, assuming the Standard Solar Model is used for the initial flux \cite{Bahcall:2004qv,survival}.  In our model, however, a large number of these low energy neutrinos could potentially oscillate into sterile neutrinos, due to the small mass squared splitting of the MaVaN pair and a potential level crossing in the sun from the MSW effect.

In order to check the acceptability of our scenario for solar neutrinos, we compute the effective mass eigenstates and mixing parameters, as a function of the electron density, for the following neutrino mass matrix. 

\be
m_D=\begin{pmatrix}0.08& 5&5\\-0.04& 6& 40\\0.04&-5& 40\end{pmatrix} \meV
 ,\quad k\vev{\acc}= \begin{pmatrix}0.1&2\times 10^{-5}&10^{-4}\\2 \times 10^{-5}&  3\times 10^{-9}& 10^{-9}\\ 10^{-5}&10^{-9}& 10^{-9}\end{pmatrix} \meV
\ee
 The Dirac mass matrix has been chosen to produce a mixing matrix and  eigenvalues consistent with the usual  interpretation of neutrino experiments \cite{review}
and the matrix $k\acc$ was chosen to be consistent with our dark energy model,  the constraints discussed above, and a Frogatt-Nielson \cite{fn} explanation of the hierarchy between the couplings of $N_1$ and $N_{2,3}$ evident in this matrix.

With these matrices, the heavy pseudo Dirac pair has mass 0.056 eV and mass squared splitting of
 $3.6\times 10^{-13} \ev^2$, the second pair has mass 0.0092 eV and mass squared splitting of  $1.7\times 10^{-13} \ev^2$ and the third pair has masses of 0.00016 eV and 0.00006 eV.
In all neutrino experiments  except   solar, the oscillation lengths associated with the sterile neutrinos are too long for such oscillations to be observable, and the oscillations among active neutrinos are characterized by the mixing matrix
\be
V_\nu=\begin{pmatrix}0.84&0.53&-0.09\\
-0.43&0.56&-0.71\\
0.33&-0.63&-0.70\end{pmatrix}\ ,\ee
where the rows correspond to flavors $e,\mu,\tau$ and columns to masses from lightest to heaviest.

A full treatment of the effects of the sterile neutrinos on solar neutrino experiments is beyond the scope of the paper, but we can make a crude  estimate of the effects as follows.

Flavor evolution of solar neutrinos propagating outward from the core of the sun may be computed using the following effective six state Hamiltonian:

\be\label{effectiveh}
H_{\rm eff}=M^2/2E + V_{\rm cc} + V_{\rm nc}
\ee 
where $M$ is the full six by six Majorana neutrino mass matrix,   $V_{\rm cc}$ raises the electron neutrino energy by an amount $\sqrt2 G_F \rho_e$,  $V_{\rm nc}$ lowers all the active neutrino energies by an amount $\frac{G_F }{\sqrt2}\rho_n$, and $\rho_e$ and $\rho_n$ are the electron and neutron densities respectively \cite{wolfenstein,Mikheev:1986gs}.
In figure 1 we plot the eigenvalues of $2 E H_{\rm eff}$, as a function of $x\equiv 2 \sqrt2 EG_F \rho_e$, with $\rho_n$ fixed to $\rho_e/2$.  As in the usual large angle MSW solution to the solar neutrino problem,  at  $x\sim 10^{-4}\eV^2$, a typical value for ${}^8B$ neutrinos produced in the core of the sun, the third heaviest eigenstate is mostly electron neutrino, and this state adiabatically evolves into into a linear combination of $e,\mu,\tau$ as it exits the sun. The linear combination is not quite a vacuum eigenstate, due to the tiny splitting of the pseudo-Dirac pair, but for neutrinos with energies above 0.02 MeV the oscillation length is longer than an AU. Thus  this state undergoes a nonadiabatic transition to the active  linear combination of the third and fourth states, and  arrives in roughly equal proportions as  $e$,  $\mu$ and $\tau$, as is consistent with the $SNO$ and $SuperK$ experiments.

The Gallium experiments are sensitive to neutrinos with energies as low as $0.2$ MeV, which are predominantly produced at $x\sim 10^{-6}\eV^2$. These neutrinos are produced in a linear combination of the third and fifth eigenstates of $H_{\rm eff}$.  For large mixing angles and mass squared splitting above $\sim 10^{-8}\ev^2$, the solar evolution   is adiabatic \cite{friedland}. As the neutrinos on the third  branch exit the sun,  they   undergo a non adiabatic transition  into the active superposition of the nearly degenerate third and fourth eigenstates. Since  the neutrinos on the fifth branch adiabatically evolve into the fifth  vacuum eigenstate, the electron neutrino survival probability is only 0.23, about 40\% 
the  LMA value. 

These results show   the model can easily reproduce the results of the SNO \cite{sno},  SuperK \cite{super}  and Homestake \cite{chlorine} experiments, but the   parameters are likely to rather constrained  in order for the survival probability of lower energy electron neutrinos to be as high as indicated by the  Gallium experiments \cite{Abdurashitov:1999zd,Hampel:1998xg,Altmann:2000ft}. The basic problem is that in the vacuum, the lightest eigenstate is mostly electron neutrino. If this neutrino   undergoes an adiabatic level crossing  in the sun with the lightest sterile neutrino, too few low energy electron neutrinos will be seen. One way to avoid this problem is to change the MaVaN parameters so that the transition   is non adiabatic and suppressed.  Either the mixing angle or the mass-squared difference can be reduced. In the next section we discuss a more interesting way to achieve experimental agreement:   addition of an acceleron coupling to matter can  raise the mass of the sterile neutrino in the sun and avoid  the level crossing.
\begin{figure}
\vspace* {0.0in}
\epsfysize=3.0in
\centerline{\epsfbox{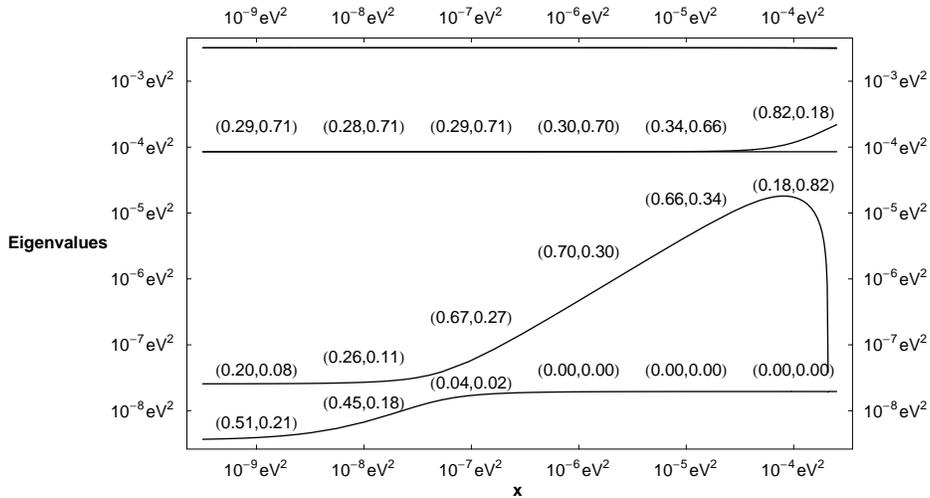}}
\caption{Plot of the eigenvalues of the matrix $2 E H_{\rm eff}$ as a function of
$x\equiv 2 \sqrt2 G_F \rho_e E_\nu$ for the parameters described in section \ref{sec:solar}. At several values of $x$, the compositions of the third, fifth and sixth eigenvalues are given  as $(P_e, P_\mu+P_\tau)$, where $P_{e,\mu,\tau}$ is the probability that this eigenstate will be detected with flavor $e,\mu,\tau$ respectively. The heaviest two eigenstates are a pseudo-Dirac pair whose separation cannot be discerned on this plot, and which have negligible electron component. 
The fourth eigenstate is nearly degenerate with the third over most of this region, and is mostly sterile.
}
\label{fig1}
\end{figure}

\subsection{Solar Neutrino Oscillations with matter dependent Acceleron expectation value}
\label{sec:solaccel}
In a previous paper, it was noted that a gravitational strength coupling of the acceleron to  electrons and nucleons could have dramatic consequences for neutrino oscillations \cite{us2}. Here we consider the effects on neutrino experiments of the coupling
\be\label{mavanterm}
\CL \supset  \lambda_N \acc \frac{m_N}{\mpl} \bar\psi_N \psi_N \ ,
\ee
where $\psi_N$ is the nucleon field, $m_N$ is the nucleon mass, $\mpl$ is the Planck scale,  and the above is to be regarded as a term in an effective Lagrangian for energies below the QCD confinement scale. The acceleron Compton wavelength is of order 2 meters, and in  larger objects, the acceleron expectation value may be found by minimizing the effective potential
\be
V_{\rm eff} = n_\nu E_\nu(\acc)+ V_0(\acc) - \lambda_N \frac{\acc}{\mpl} \rho \ee
where $\rho$ is the mass density of the object. The second term pushes the acceleron towards zero while the first and third terms tend to increase the acceleron expectation value. In space, the first term is more important, while for  matter at densities above $\sim 10^{-6} \frac{g}{{\rm cm}^3}/\lambda_N$, the second term is more important, and the acceleron expectation value increases. Because of the small coupling of the acceleron to the heavier neutrinos,  for $\rho \lambda_N \lesssim 10^5 \frac{g}{{\rm cm}^3}$, such a change in the acceleron expectation value has negligible effect on the two heavier pseudo Dirac pairs. The two lighter eigenvalues, however are very sensitive to such a change. 

For $10^{-8}\lesssim\lambda_N\lesssim 10^3$, the main effect of the coupling \eref{mavanterm} on the evolution of solar neutrinos  is to change the eigenvalues and flavor composition of the  two lightest eigenstates of the effective hamiltonian \ref{effectiveh}. In figures 2-4  we show the eigenvalues of $H_{\rm eff}$, as a function of $y \equiv   2    \sqrt2  G_F \rho_e$  for three different neutrino energies and  $\rho_n$ fixed to $\rho_e/2$. We have taken the acceleron vev to be
$\vev{\acc}= \vev{\acc_0}[1+ 10^7 y (MeV/eV^2)]$, a crude approximation to the expected profile for
$\lambda_N\sim 10^{-5}$.   

\begin{figure}
\vspace* {0.0in}
\epsfysize=3.0in
\centerline{\epsfbox{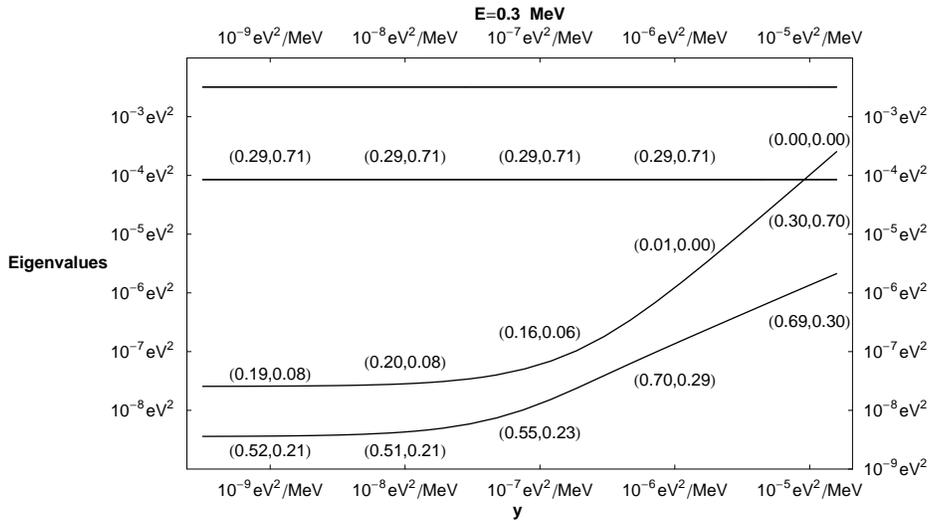}}
\caption{Plot of the eigenvalues of the matrix $2 E H_{\rm eff}$ as a function of
$y\equiv 2 \sqrt2 G_F \rho_e $, for the scenario described in section \ref{sec:solaccel} and an energy of 0.3 MeV, which is typical for the Gallium experiments. At several values of $y$, the compositions of the third, fifth and sixth eigenvalues are given  as $(P_e, P_\mu+P_\tau)$, where $P_{e,\mu,\tau}$ is the probability that this eigenstate will be detected with flavor $e,\mu,\tau$ respectively.
}
\label{fig2}
\end{figure}
\begin{figure}
\vspace* {0.0in}
\epsfysize=3.0in
\centerline{\epsfbox{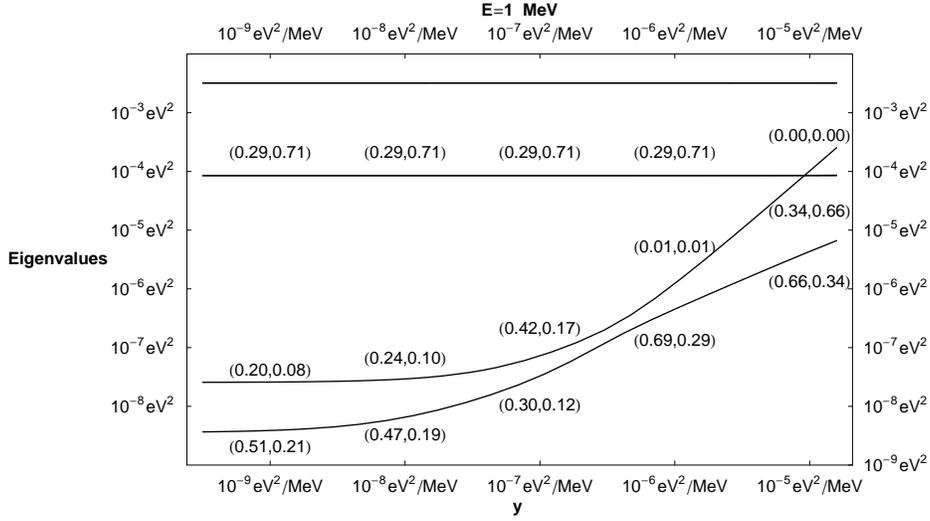}}
\caption{Plot of the eigenvalues of the matrix $2 E H_{\rm eff}$ as a function of
$y $, for the scenario described in section \ref{sec:solaccel} and an energy of 1 MeV, which is typical for the Homestake experiment. At several values of $y$, the compositions of the third, fifth and sixth eigenvalues are given.}
\label{fig3}
\end{figure}
\begin{figure}
\vspace* {0.0in}
\epsfysize=3.0in
\centerline{\epsfbox{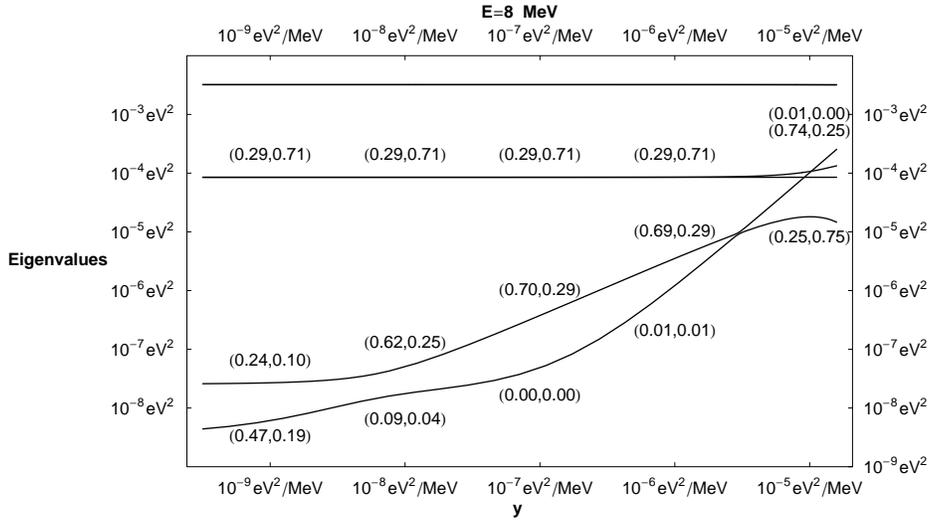}}
\caption{Plot of the eigenvalues of the matrix $2 E H_{\rm eff}$ as a function of
$y$,  for the scenario described in section \ref{sec:solaccel} and an 8 MeV energy, which is typical for the SuperKamionkande and SNO experiments. At several values of $y$, the compositions of the third, fifth and sixth eigenvalues are given.
}
\label{fig4}
\end{figure}

We see that  the main effect of the inclusion of the term \eref{mavanterm} is raise the mass of the lightest sterile state   in the sun, and thus to prevent the level crossing at $x\sim 10^{-8} \eV^2$ between an active and sterile neutrino. At low energies  the lightest eigenstate of $H_{\rm eff}$ inside the sun is  then mainly active. If neutrinos on this branch evolve adiabatically they exit the sun as the lightest vacuum  eigenstate which is also mostly active. One can thus easily achieve a sufficiently high survival probability for low energy electron neutrinos. For optimal choice of the coupling $\lambda_N$, one could still have the level crossing with the sterile neutrino for intermediate energy neutrinos.  This could potentially achieve better agreement with experiment than the LMA MSW parameters\cite{deHolanda:2002ma}, by substantially suppressing the survival of electron neutrinos in the $1-6$ MeV   range, retaining the higher survival probability of lower energy neutrinos, and leaving the survival of higher energy neutrinos unaffected.

Although we have specialized our discussion to case 1, note that the novel solar neutrino phenomenology of the light MaVaN pair of neutrinos is identical for both cases.

 \subsection{New Forces}
 For non-zero $\lambda_N$ the acceleron will mediate a   new force, which can be searched for in tests of gravity at distance scales shorter than 2 meters 
\cite{Adelberger:1992ph,gravityreviews}. In order to compute the experimental constraints on $\lambda_N$, the effects of  the nonlinear  terms in the acceleron  equation of motion  need  to be considered, as these can dramatically weaken the constraints \cite{chameleon,Gubser:2004uf,feldman}.

\subsection{Double Beta Decay }
The consequences of this neutrino mass model for  neutrinoless double beta decay are simple: there should be no  observable neutrinoless decays, even though our neutrinos are Majorana. This would be the case even in an inverted hierarchy or degenerate neutrino scenario. The explanation is simplest if one works in the flavor basis. All the entries in our six by six dimensional neutrino mass matrix are much smaller then the MeV energies typical of beta decay. Therefore in computing  the double beta decay rate the mass insertion approximation is excellent. In this approximation the rate is proportional to the $m_{ee}$  component of the neutrino mass matrix. This component is zero in the seesaw model. The result that neutrinoless double beta decay is unobservable  is radically different from   the  conventional  seesaw model, where one assumes the sterile neutrinos are much  heavier than the beta decay energy scale. 

Double beta decay with acceleron or other light scalar emission is allowed but is also too suppressed to observe, as only the sterile neutrinos couple directly to light scalars. Thus such processes require two insertions of the Dirac mass term, in order to connect the active and sterile neutrinos, as well as the Yukawa coupling of the sterile neutrino to the scalars.

\section{Summary and Conclusions}
\label{sec:conclusions}
The MaVaN dark energy model offers the prospect of testable interactions between the dark energy sector and neutrinos. However our previous work did not address the issues of the origin of the acceleron potential, the degree of fine tuning required, or the stability of the dark energy fluid against growth of inhomogeneities.
We have shown that in a straightforward supersymmetrization of the MaVaN dark energy model, the scalar partner of a sterile neutrino acts like the waterfall field of hybrid inflation, whose potential energy can drive accelerated expansion of the universe.  Supersymmetry can stabilize the required mass scales and couplings against fine-tuning, provided that the MaVaN is is lighter than of order $10^{-4}$ eV. A $10^{-4}$ eV MaVaN is sufficiently relativistic to prevent inhomogeneous growth of neutrino density.
We show that a model with three active and three sterile neutrinos can satisfy all experimental constraints while providing a natural theory of dark energy. Possible signatures of the model include a reduction of the electron neutrino survival probability of low and intermediate energy solar neutrinos, long baseline oscillations of muon and tau neutrinos into sterile neutrinos, suppression of neutrinoless double beta decay, and novel dependence of neutrino oscillation parameters on the environment.

As this paper was being completed ref.\cite{Takahashi:2005kw} appeared, which presents a somewhat different supersymmetric model of MaVaN dark energy.

\section*{Acknowledgements} The work of   A. Nelson was partially
supported by the DOE under contract DE-FGO3-96-ER40956, and a Guggenheim fellowship. N. Weiner was supported by NSF CAREER grant PHY-0449818. We thank Alex Friedland, David Kaplan, Cecilia Lunardini and Kathryn Zurek for useful conversations.
\bibliography{hybrid}

\begin{thebibliography}{58}
\expandafter\ifx\csname natexlab\endcsname\relax\def\natexlab#1{#1}\fi
\expandafter\ifx\csname bibnamefont\endcsname\relax
  \def\bibnamefont#1{#1}\fi
\expandafter\ifx\csname bibfnamefont\endcsname\relax
  \def\bibfnamefont#1{#1}\fi
\expandafter\ifx\csname citenamefont\endcsname\relax
  \def\citenamefont#1{#1}\fi
\expandafter\ifx\csname url\endcsname\relax
  \def\url#1{\texttt{#1}}\fi
\expandafter\ifx\csname urlprefix\endcsname\relax\def\urlprefix{URL }\fi
\providecommand{\bibinfo}[2]{#2}
\providecommand{\eprint}[2][]{\url{#2}}

\bibitem[{\citenamefont{Abazajian et~al.}(2004)}]{Abazajian:2004aj}
\bibinfo{author}{\bibfnamefont{K.}~\bibnamefont{Abazajian}}
  \bibnamefont{et~al.} (\bibinfo{collaboration}{SDSS}),
  \bibinfo{journal}{Astron. J.} \textbf{\bibinfo{volume}{128}},
  \bibinfo{pages}{502} (\bibinfo{year}{2004}), \eprint{astro-ph/0403325}.

\bibitem[{\citenamefont{Spergel et~al.}(2003)}]{Spergel:2003cb}
\bibinfo{author}{\bibfnamefont{D.~N.} \bibnamefont{Spergel}}
  \bibnamefont{et~al.} (\bibinfo{collaboration}{WMAP}),
  \bibinfo{journal}{Astrophys. J. Suppl.} \textbf{\bibinfo{volume}{148}},
  \bibinfo{pages}{175} (\bibinfo{year}{2003}), \eprint{astro-ph/0302209}.

\bibitem[{\citenamefont{Percival et~al.}(2001)}]{Percival:2001hw}
\bibinfo{author}{\bibfnamefont{W.~J.} \bibnamefont{Percival}}
  \bibnamefont{et~al.} (\bibinfo{collaboration}{The 2dFGRS}),
  \bibinfo{journal}{Mon. Not. Roy. Astron. Soc.}
  \textbf{\bibinfo{volume}{327}}, \bibinfo{pages}{1297} (\bibinfo{year}{2001}),
  \eprint{astro-ph/0105252}.

\bibitem[{\citenamefont{de~Bernardis et~al.}(2000)}]{deBernardis:2000gy}
\bibinfo{author}{\bibfnamefont{P.}~\bibnamefont{de~Bernardis}}
  \bibnamefont{et~al.} (\bibinfo{collaboration}{Boomerang}),
  \bibinfo{journal}{Nature} \textbf{\bibinfo{volume}{404}},
  \bibinfo{pages}{955} (\bibinfo{year}{2000}), \eprint{astro-ph/0004404}.

\bibitem[{\citenamefont{Lee et~al.}(2001)}]{Lee:2001yp}
\bibinfo{author}{\bibfnamefont{A.~T.} \bibnamefont{Lee}} \bibnamefont{et~al.},
  \bibinfo{journal}{Astrophys. J.} \textbf{\bibinfo{volume}{561}},
  \bibinfo{pages}{L1} (\bibinfo{year}{2001}), \eprint{astro-ph/0104459}.

\bibitem[{\citenamefont{Halverson et~al.}(2002)}]{Halverson:2001yy}
\bibinfo{author}{\bibfnamefont{N.~W.} \bibnamefont{Halverson}}
  \bibnamefont{et~al.}, \bibinfo{journal}{Astrophys. J.}
  \textbf{\bibinfo{volume}{568}}, \bibinfo{pages}{38} (\bibinfo{year}{2002}),
  \eprint{astro-ph/0104489}.

\bibitem[{\citenamefont{Riess et~al.}(1998)}]{Riess:1998cb}
\bibinfo{author}{\bibfnamefont{A.~G.} \bibnamefont{Riess}} \bibnamefont{et~al.}
  (\bibinfo{collaboration}{Supernova Search Team}), \bibinfo{journal}{Astron.
  J.} \textbf{\bibinfo{volume}{116}}, \bibinfo{pages}{1009}
  (\bibinfo{year}{1998}), \eprint{astro-ph/9805201}.

\bibitem[{\citenamefont{Perlmutter et~al.}(1999)}]{Perlmutter:1998np}
\bibinfo{author}{\bibfnamefont{S.}~\bibnamefont{Perlmutter}}
  \bibnamefont{et~al.} (\bibinfo{collaboration}{Supernova Cosmology Project}),
  \bibinfo{journal}{Astrophys. J.} \textbf{\bibinfo{volume}{517}},
  \bibinfo{pages}{565} (\bibinfo{year}{1999}), \eprint{astro-ph/9812133}.

\bibitem[{\citenamefont{Fardon et~al.}(2004)\citenamefont{Fardon, Nelson, and
  Weiner}}]{us}
\bibinfo{author}{\bibfnamefont{R.}~\bibnamefont{Fardon}},
  \bibinfo{author}{\bibfnamefont{A.~E.} \bibnamefont{Nelson}},
  \bibnamefont{and} \bibinfo{author}{\bibfnamefont{N.}~\bibnamefont{Weiner}},
  \bibinfo{journal}{JCAP} \textbf{\bibinfo{volume}{0410}}, \bibinfo{pages}{005}
  (\bibinfo{year}{2004}), \eprint{astro-ph/0309800}.

\bibitem[{\citenamefont{Wolfenstein}(1978)}]{wolfenstein}
\bibinfo{author}{\bibfnamefont{L.}~\bibnamefont{Wolfenstein}},
  \bibinfo{journal}{Phys. Rev.} \textbf{\bibinfo{volume}{D17}},
  \bibinfo{pages}{2369} (\bibinfo{year}{1978}).

\bibitem[{\citenamefont{Sawyer}(1999)}]{Sawyer:1998ac}
\bibinfo{author}{\bibfnamefont{R.~F.} \bibnamefont{Sawyer}},
  \bibinfo{journal}{Phys. Lett.} \textbf{\bibinfo{volume}{B448}},
  \bibinfo{pages}{174} (\bibinfo{year}{1999}), \eprint{hep-ph/9809348}.

\bibitem[{\citenamefont{Kawasaki et~al.}(1992)\citenamefont{Kawasaki, Murayama,
  and Yanagida}}]{Kawasaki:1991gn}
\bibinfo{author}{\bibfnamefont{M.}~\bibnamefont{Kawasaki}},
  \bibinfo{author}{\bibfnamefont{H.}~\bibnamefont{Murayama}}, \bibnamefont{and}
  \bibinfo{author}{\bibfnamefont{T.}~\bibnamefont{Yanagida}},
  \bibinfo{journal}{Mod. Phys. Lett.} \textbf{\bibinfo{volume}{A7}},
  \bibinfo{pages}{563} (\bibinfo{year}{1992}).

\bibitem[{\citenamefont{Stephenson et~al.}(1998)\citenamefont{Stephenson,
  Goldman, and McKellar}}]{Stephenson:1996qj}
\bibinfo{author}{\bibfnamefont{J.}~\bibnamefont{Stephenson},
  \bibfnamefont{G.~J.}},
  \bibinfo{author}{\bibfnamefont{T.}~\bibnamefont{Goldman}}, \bibnamefont{and}
  \bibinfo{author}{\bibfnamefont{B.~H.~J.} \bibnamefont{McKellar}},
  \bibinfo{journal}{Int. J. Mod. Phys.} \textbf{\bibinfo{volume}{A13}},
  \bibinfo{pages}{2765} (\bibinfo{year}{1998}), \eprint{hep-ph/9603392}.

\bibitem[{\citenamefont{Kaplan et~al.}(2004)\citenamefont{Kaplan, Nelson, and
  Weiner}}]{us2}
\bibinfo{author}{\bibfnamefont{D.~B.} \bibnamefont{Kaplan}},
  \bibinfo{author}{\bibfnamefont{A.~E.} \bibnamefont{Nelson}},
  \bibnamefont{and} \bibinfo{author}{\bibfnamefont{N.}~\bibnamefont{Weiner}},
  \bibinfo{journal}{Phys. Rev. Lett.} \textbf{\bibinfo{volume}{93}},
  \bibinfo{pages}{091801} (\bibinfo{year}{2004}), \eprint{hep-ph/0401099}.

\bibitem[{\citenamefont{Bi et~al.}(2004{\natexlab{a}})\citenamefont{Bi, Gu,
  Wang, and Zhang}}]{Bi:2003yr}
\bibinfo{author}{\bibfnamefont{X.-J.} \bibnamefont{Bi}},
  \bibinfo{author}{\bibfnamefont{P.-h.} \bibnamefont{Gu}},
  \bibinfo{author}{\bibfnamefont{X.-l.} \bibnamefont{Wang}}, \bibnamefont{and}
  \bibinfo{author}{\bibfnamefont{X.-m.} \bibnamefont{Zhang}},
  \bibinfo{journal}{Phys. Rev.} \textbf{\bibinfo{volume}{D69}},
  \bibinfo{pages}{113007} (\bibinfo{year}{2004}{\natexlab{a}}),
  \eprint{hep-ph/0311022}.

\bibitem[{\citenamefont{Hung and Pas}(2005)}]{Hung:2003jb}
\bibinfo{author}{\bibfnamefont{P.~Q.} \bibnamefont{Hung}} \bibnamefont{and}
  \bibinfo{author}{\bibfnamefont{H.}~\bibnamefont{Pas}}, \bibinfo{journal}{Mod.
  Phys. Lett.} \textbf{\bibinfo{volume}{A20}}, \bibinfo{pages}{1209}
  (\bibinfo{year}{2005}), \eprint{astro-ph/0311131}.

\bibitem[{\citenamefont{Li et~al.}(2005)\citenamefont{Li, Dai, and
  Zhang}}]{Li:2004tq}
\bibinfo{author}{\bibfnamefont{H.}~\bibnamefont{Li}},
  \bibinfo{author}{\bibfnamefont{Z.-g.} \bibnamefont{Dai}}, \bibnamefont{and}
  \bibinfo{author}{\bibfnamefont{X.-m.} \bibnamefont{Zhang}},
  \bibinfo{journal}{Phys. Rev.} \textbf{\bibinfo{volume}{D71}},
  \bibinfo{pages}{113003} (\bibinfo{year}{2005}), \eprint{hep-ph/0411228}.

\bibitem[{\citenamefont{Barger et~al.}(2005{\natexlab{a}})\citenamefont{Barger,
  Huber, and Marfatia}}]{Barger:2005mn}
\bibinfo{author}{\bibfnamefont{V.}~\bibnamefont{Barger}},
  \bibinfo{author}{\bibfnamefont{P.}~\bibnamefont{Huber}}, \bibnamefont{and}
  \bibinfo{author}{\bibfnamefont{D.}~\bibnamefont{Marfatia}}
  (\bibinfo{year}{2005}{\natexlab{a}}), \eprint{hep-ph/0502196}.

\bibitem[{\citenamefont{Cirelli et~al.}(2005)\citenamefont{Cirelli,
  Gonzalez-Garcia, and Pena-Garay}}]{Cirelli:2005sg}
\bibinfo{author}{\bibfnamefont{M.}~\bibnamefont{Cirelli}},
  \bibinfo{author}{\bibfnamefont{M.~C.} \bibnamefont{Gonzalez-Garcia}},
  \bibnamefont{and}
  \bibinfo{author}{\bibfnamefont{C.}~\bibnamefont{Pena-Garay}},
  \bibinfo{journal}{Nucl. Phys.} \textbf{\bibinfo{volume}{B719}},
  \bibinfo{pages}{219} (\bibinfo{year}{2005}), \eprint{hep-ph/0503028}.

\bibitem[{\citenamefont{Bi et~al.}(2004{\natexlab{b}})\citenamefont{Bi, Feng,
  Li, and Zhang}}]{Bi:2004ns}
\bibinfo{author}{\bibfnamefont{X.-J.} \bibnamefont{Bi}},
  \bibinfo{author}{\bibfnamefont{B.}~\bibnamefont{Feng}},
  \bibinfo{author}{\bibfnamefont{H.}~\bibnamefont{Li}}, \bibnamefont{and}
  \bibinfo{author}{\bibfnamefont{X.-m.} \bibnamefont{Zhang}}
  (\bibinfo{year}{2004}{\natexlab{b}}), \eprint{hep-ph/0412002}.

\bibitem[{\citenamefont{Brookfield et~al.}(2005)\citenamefont{Brookfield,
  van~de Bruck, Mota, and Tocchini-Valentini}}]{Brookfield:2005td}
\bibinfo{author}{\bibfnamefont{A.~W.} \bibnamefont{Brookfield}},
  \bibinfo{author}{\bibfnamefont{C.}~\bibnamefont{van~de Bruck}},
  \bibinfo{author}{\bibfnamefont{D.~F.} \bibnamefont{Mota}}, \bibnamefont{and}
  \bibinfo{author}{\bibfnamefont{D.}~\bibnamefont{Tocchini-Valentini}}
  (\bibinfo{year}{2005}), \eprint{astro-ph/0503349}.

\bibitem[{\citenamefont{Peccei}(2005)}]{Peccei:2004sz}
\bibinfo{author}{\bibfnamefont{R.~D.} \bibnamefont{Peccei}},
  \bibinfo{journal}{Phys. Rev.} \textbf{\bibinfo{volume}{D71}},
  \bibinfo{pages}{023527} (\bibinfo{year}{2005}), \eprint{hep-ph/0411137}.

\bibitem[{\citenamefont{Hill and Ross}(1988)}]{Hill:1988bu}
\bibinfo{author}{\bibfnamefont{C.~T.} \bibnamefont{Hill}} \bibnamefont{and}
  \bibinfo{author}{\bibfnamefont{G.~G.} \bibnamefont{Ross}},
  \bibinfo{journal}{Nucl. Phys.} \textbf{\bibinfo{volume}{B311}},
  \bibinfo{pages}{253} (\bibinfo{year}{1988}).

\bibitem[{\citenamefont{Hung}(2000)}]{Hung:2000yg}
\bibinfo{author}{\bibfnamefont{P.~Q.} \bibnamefont{Hung}}
  (\bibinfo{year}{2000}), \eprint{hep-ph/0010126}.

\bibitem[{\citenamefont{Barbieri et~al.}(2005)\citenamefont{Barbieri, Hall,
  Oliver, and Strumia}}]{Barbieri:2005gj}
\bibinfo{author}{\bibfnamefont{R.}~\bibnamefont{Barbieri}},
  \bibinfo{author}{\bibfnamefont{L.~J.} \bibnamefont{Hall}},
  \bibinfo{author}{\bibfnamefont{S.~J.} \bibnamefont{Oliver}},
  \bibnamefont{and} \bibinfo{author}{\bibfnamefont{A.}~\bibnamefont{Strumia}}
  (\bibinfo{year}{2005}), \eprint{hep-ph/0505124}.

\bibitem[{\citenamefont{Guendelman and Kaganovich}(2004)}]{Guendelman:2004hk}
\bibinfo{author}{\bibfnamefont{E.~I.} \bibnamefont{Guendelman}}
  \bibnamefont{and} \bibinfo{author}{\bibfnamefont{A.~B.}
  \bibnamefont{Kaganovich}} (\bibinfo{year}{2004}), \eprint{hep-th/0411188}.

\bibitem[{\citenamefont{Singh}(1995)}]{Singh:1994nt}
\bibinfo{author}{\bibfnamefont{A.}~\bibnamefont{Singh}},
  \bibinfo{journal}{Phys. Rev.} \textbf{\bibinfo{volume}{D52}},
  \bibinfo{pages}{6700} (\bibinfo{year}{1995}), \eprint{hep-ph/9412240}.

\bibitem[{\citenamefont{Linde}(1994)}]{linde}
\bibinfo{author}{\bibfnamefont{A.~D.} \bibnamefont{Linde}},
  \bibinfo{journal}{Phys. Rev.} \textbf{\bibinfo{volume}{D49}},
  \bibinfo{pages}{748} (\bibinfo{year}{1994}), \eprint{astro-ph/9307002}.

\bibitem[{\citenamefont{Afshordi et~al.}(2005)\citenamefont{Afshordi,
  Zaldarriaga, and Kohri}}]{matias}
\bibinfo{author}{\bibfnamefont{N.}~\bibnamefont{Afshordi}},
  \bibinfo{author}{\bibfnamefont{M.}~\bibnamefont{Zaldarriaga}},
  \bibnamefont{and} \bibinfo{author}{\bibfnamefont{K.}~\bibnamefont{Kohri}}
  (\bibinfo{year}{2005}), \eprint{astro-ph/0506663}.

\bibitem[{\citenamefont{Dine and Nelson}(1993)}]{Dine:1993yw}
\bibinfo{author}{\bibfnamefont{M.}~\bibnamefont{Dine}} \bibnamefont{and}
  \bibinfo{author}{\bibfnamefont{A.~E.} \bibnamefont{Nelson}},
  \bibinfo{journal}{Phys. Rev.} \textbf{\bibinfo{volume}{D48}},
  \bibinfo{pages}{1277} (\bibinfo{year}{1993}), \eprint{hep-ph/9303230}.

\bibitem[{\citenamefont{Dine et~al.}(1995)\citenamefont{Dine, Nelson, and
  Shirman}}]{Dine:1994vc}
\bibinfo{author}{\bibfnamefont{M.}~\bibnamefont{Dine}},
  \bibinfo{author}{\bibfnamefont{A.~E.} \bibnamefont{Nelson}},
  \bibnamefont{and} \bibinfo{author}{\bibfnamefont{Y.}~\bibnamefont{Shirman}},
  \bibinfo{journal}{Phys. Rev.} \textbf{\bibinfo{volume}{D51}},
  \bibinfo{pages}{1362} (\bibinfo{year}{1995}), \eprint{hep-ph/9408384}.

\bibitem[{\citenamefont{Dine et~al.}(1996)\citenamefont{Dine, Nelson, Nir, and
  Shirman}}]{Dine:1995ag}
\bibinfo{author}{\bibfnamefont{M.}~\bibnamefont{Dine}},
  \bibinfo{author}{\bibfnamefont{A.~E.} \bibnamefont{Nelson}},
  \bibinfo{author}{\bibfnamefont{Y.}~\bibnamefont{Nir}}, \bibnamefont{and}
  \bibinfo{author}{\bibfnamefont{Y.}~\bibnamefont{Shirman}},
  \bibinfo{journal}{Phys. Rev.} \textbf{\bibinfo{volume}{D53}},
  \bibinfo{pages}{2658} (\bibinfo{year}{1996}), \eprint{hep-ph/9507378}.

\bibitem[{\citenamefont{Giudice and Rattazzi}(1999)}]{Giudice:1998bp}
\bibinfo{author}{\bibfnamefont{G.~F.} \bibnamefont{Giudice}} \bibnamefont{and}
  \bibinfo{author}{\bibfnamefont{R.}~\bibnamefont{Rattazzi}},
  \bibinfo{journal}{Phys. Rept.} \textbf{\bibinfo{volume}{322}},
  \bibinfo{pages}{419} (\bibinfo{year}{1999}), \eprint{hep-ph/9801271}.

\bibitem[{\citenamefont{de~Gouvea et~al.}(1997)\citenamefont{de~Gouvea, Moroi,
  and Murayama}}]{deGouvea:1997tn}
\bibinfo{author}{\bibfnamefont{A.}~\bibnamefont{de~Gouvea}},
  \bibinfo{author}{\bibfnamefont{T.}~\bibnamefont{Moroi}}, \bibnamefont{and}
  \bibinfo{author}{\bibfnamefont{H.}~\bibnamefont{Murayama}},
  \bibinfo{journal}{Phys. Rev.} \textbf{\bibinfo{volume}{D56}},
  \bibinfo{pages}{1281} (\bibinfo{year}{1997}), \eprint{hep-ph/9701244}.

\bibitem[{\citenamefont{Randall and Sundrum}(1999)}]{Randall:1998uk}
\bibinfo{author}{\bibfnamefont{L.}~\bibnamefont{Randall}} \bibnamefont{and}
  \bibinfo{author}{\bibfnamefont{R.}~\bibnamefont{Sundrum}},
  \bibinfo{journal}{Nucl. Phys.} \textbf{\bibinfo{volume}{B557}},
  \bibinfo{pages}{79} (\bibinfo{year}{1999}), \eprint{hep-th/9810155}.

\bibitem[{\citenamefont{Arkani-Hamed and Schmaltz}(2000)}]{Arkani-Hamed:1999dc}
\bibinfo{author}{\bibfnamefont{N.}~\bibnamefont{Arkani-Hamed}}
  \bibnamefont{and} \bibinfo{author}{\bibfnamefont{M.}~\bibnamefont{Schmaltz}},
  \bibinfo{journal}{Phys. Rev.} \textbf{\bibinfo{volume}{D61}},
  \bibinfo{pages}{033005} (\bibinfo{year}{2000}), \eprint{hep-ph/9903417}.

\bibitem[{\citenamefont{Weiner and Zurek}(2005)}]{Weiner:2005ac}
\bibinfo{author}{\bibfnamefont{N.}~\bibnamefont{Weiner}} \bibnamefont{and}
  \bibinfo{author}{\bibfnamefont{K.}~\bibnamefont{Zurek}}
  (\bibinfo{year}{2005}), \eprint{hep-ph/0509201}.

\bibitem[{\citenamefont{Zurek}(2004)}]{Zurek:2004vd}
\bibinfo{author}{\bibfnamefont{K.~M.} \bibnamefont{Zurek}},
  \bibinfo{journal}{JHEP} \textbf{\bibinfo{volume}{10}}, \bibinfo{pages}{058}
  (\bibinfo{year}{2004}), \eprint{hep-ph/0405141}.

\bibitem[{\citenamefont{Athanassopoulos et~al.}(1998)}]{Athanassopoulos:1998pv}
\bibinfo{author}{\bibfnamefont{C.}~\bibnamefont{Athanassopoulos}}
  \bibnamefont{et~al.} (\bibinfo{collaboration}{LSND}), \bibinfo{journal}{Phys.
  Rev. Lett.} \textbf{\bibinfo{volume}{81}}, \bibinfo{pages}{1774}
  (\bibinfo{year}{1998}), \eprint{nucl-ex/9709006}.

\bibitem[{\citenamefont{Ahmed et~al.}(2003)}]{sno}
\bibinfo{author}{\bibfnamefont{S.~N.} \bibnamefont{Ahmed}} \bibnamefont{et~al.}
  (\bibinfo{collaboration}{SNO}) (\bibinfo{year}{2003}),
  \eprint{nucl-ex/0309004}.

\bibitem[{\citenamefont{Miura}(2002)}]{super}
\bibinfo{author}{\bibfnamefont{M.}~\bibnamefont{Miura}},
  \bibinfo{journal}{Nucl. Phys. Proc. Suppl.} \textbf{\bibinfo{volume}{111}},
  \bibinfo{pages}{158} (\bibinfo{year}{2002}).

\bibitem[{\citenamefont{Mikheev and Smirnov}(1985)}]{Mikheev:1986gs}
\bibinfo{author}{\bibfnamefont{S.~P.} \bibnamefont{Mikheev}} \bibnamefont{and}
  \bibinfo{author}{\bibfnamefont{A.~Y.} \bibnamefont{Smirnov}},
  \bibinfo{journal}{Sov. J. Nucl. Phys.} \textbf{\bibinfo{volume}{42}},
  \bibinfo{pages}{913} (\bibinfo{year}{1985}).

\bibitem[{\citenamefont{Abdurashitov et~al.}(1999)}]{Abdurashitov:1999zd}
\bibinfo{author}{\bibfnamefont{J.~N.} \bibnamefont{Abdurashitov}}
  \bibnamefont{et~al.} (\bibinfo{collaboration}{SAGE}), \bibinfo{journal}{Phys.
  Rev.} \textbf{\bibinfo{volume}{C60}}, \bibinfo{pages}{055801}
  (\bibinfo{year}{1999}), \eprint{astro-ph/9907113}.

\bibitem[{\citenamefont{Hampel et~al.}(1999)}]{Hampel:1998xg}
\bibinfo{author}{\bibfnamefont{W.}~\bibnamefont{Hampel}} \bibnamefont{et~al.}
  (\bibinfo{collaboration}{GALLEX}), \bibinfo{journal}{Phys. Lett.}
  \textbf{\bibinfo{volume}{B447}}, \bibinfo{pages}{127} (\bibinfo{year}{1999}).

\bibitem[{\citenamefont{Altmann et~al.}(2000)}]{Altmann:2000ft}
\bibinfo{author}{\bibfnamefont{M.}~\bibnamefont{Altmann}} \bibnamefont{et~al.}
  (\bibinfo{collaboration}{GNO}), \bibinfo{journal}{Phys. Lett.}
  \textbf{\bibinfo{volume}{B490}}, \bibinfo{pages}{16} (\bibinfo{year}{2000}),
  \eprint{hep-ex/0006034}.

\bibitem[{\citenamefont{Bahcall}(2004)}]{Bahcall:2004qv}
\bibinfo{author}{\bibfnamefont{J.~N.} \bibnamefont{Bahcall}}
  (\bibinfo{year}{2004}), \eprint{hep-ph/0412068}.

\bibitem[{\citenamefont{Barger et~al.}(2005{\natexlab{b}})\citenamefont{Barger,
  Marfatia, and Whisnant}}]{survival}
\bibinfo{author}{\bibfnamefont{V.}~\bibnamefont{Barger}},
  \bibinfo{author}{\bibfnamefont{D.}~\bibnamefont{Marfatia}}, \bibnamefont{and}
  \bibinfo{author}{\bibfnamefont{K.}~\bibnamefont{Whisnant}},
  \bibinfo{journal}{Phys. Lett.} \textbf{\bibinfo{volume}{B617}},
  \bibinfo{pages}{78} (\bibinfo{year}{2005}{\natexlab{b}}),
  \eprint{hep-ph/0501247}.

\bibitem[{\citenamefont{Gonzalez-Garcia and Nir}(2003)}]{review}
\bibinfo{author}{\bibfnamefont{M.~C.} \bibnamefont{Gonzalez-Garcia}}
  \bibnamefont{and} \bibinfo{author}{\bibfnamefont{Y.}~\bibnamefont{Nir}},
  \bibinfo{journal}{Rev. Mod. Phys.} \textbf{\bibinfo{volume}{75}},
  \bibinfo{pages}{345} (\bibinfo{year}{2003}), \eprint{hep-ph/0202058}.

\bibitem[{\citenamefont{Froggatt and Nielsen}(1979)}]{fn}
\bibinfo{author}{\bibfnamefont{C.~D.} \bibnamefont{Froggatt}} \bibnamefont{and}
  \bibinfo{author}{\bibfnamefont{H.~B.} \bibnamefont{Nielsen}},
  \bibinfo{journal}{Nucl. Phys.} \textbf{\bibinfo{volume}{B147}},
  \bibinfo{pages}{277} (\bibinfo{year}{1979}).

\bibitem[{\citenamefont{Friedland}(2001)}]{friedland}
\bibinfo{author}{\bibfnamefont{A.}~\bibnamefont{Friedland}},
  \bibinfo{journal}{Phys. Rev.} \textbf{\bibinfo{volume}{D64}},
  \bibinfo{pages}{013008} (\bibinfo{year}{2001}), \eprint{hep-ph/0010231}.

\bibitem[{\citenamefont{Cleveland et~al.}(1998)}]{chlorine}
\bibinfo{author}{\bibfnamefont{B.~T.} \bibnamefont{Cleveland}}
  \bibnamefont{et~al.}, \bibinfo{journal}{Astrophys. J.}
  \textbf{\bibinfo{volume}{496}}, \bibinfo{pages}{505} (\bibinfo{year}{1998}).

\bibitem[{\citenamefont{de~Holanda and Smirnov}(2002)}]{deHolanda:2002ma}
\bibinfo{author}{\bibfnamefont{P.~C.} \bibnamefont{de~Holanda}}
  \bibnamefont{and} \bibinfo{author}{\bibfnamefont{A.~Y.}
  \bibnamefont{Smirnov}} (\bibinfo{year}{2002}), \eprint{hep-ph/0211264}.

\bibitem[{\citenamefont{Adelberger et~al.}(1991)\citenamefont{Adelberger,
  Heckel, Stubbs, and Rogers}}]{Adelberger:1992ph}
\bibinfo{author}{\bibfnamefont{E.~G.} \bibnamefont{Adelberger}},
  \bibinfo{author}{\bibfnamefont{B.~R.} \bibnamefont{Heckel}},
  \bibinfo{author}{\bibfnamefont{C.~W.} \bibnamefont{Stubbs}},
  \bibnamefont{and} \bibinfo{author}{\bibfnamefont{W.~F.}
  \bibnamefont{Rogers}}, \bibinfo{journal}{Ann. Rev. Nucl. Part. Sci.}
  \textbf{\bibinfo{volume}{41}}, \bibinfo{pages}{269} (\bibinfo{year}{1991}).

\bibitem[{\citenamefont{Adelberger et~al.}(2003)\citenamefont{Adelberger,
  Heckel, and Nelson}}]{gravityreviews}
\bibinfo{author}{\bibfnamefont{E.~G.} \bibnamefont{Adelberger}},
  \bibinfo{author}{\bibfnamefont{B.~R.} \bibnamefont{Heckel}},
  \bibnamefont{and} \bibinfo{author}{\bibfnamefont{A.~E.} \bibnamefont{Nelson}}
  (\bibinfo{year}{2003}), \eprint{hep-ph/0307284}.

\bibitem[{\citenamefont{Khoury and Weltman}(2004)}]{chameleon}
\bibinfo{author}{\bibfnamefont{J.}~\bibnamefont{Khoury}} \bibnamefont{and}
  \bibinfo{author}{\bibfnamefont{A.}~\bibnamefont{Weltman}},
  \bibinfo{journal}{Phys. Rev. Lett.} \textbf{\bibinfo{volume}{93}},
  \bibinfo{pages}{171104} (\bibinfo{year}{2004}), \eprint{astro-ph/0309300}.

\bibitem[{\citenamefont{Gubser and Khoury}(2004)}]{Gubser:2004uf}
\bibinfo{author}{\bibfnamefont{S.~S.} \bibnamefont{Gubser}} \bibnamefont{and}
  \bibinfo{author}{\bibfnamefont{J.}~\bibnamefont{Khoury}},
  \bibinfo{journal}{Phys. Rev.} \textbf{\bibinfo{volume}{D70}},
  \bibinfo{pages}{104001} (\bibinfo{year}{2004}), \eprint{hep-ph/0405231}.

\bibitem[{\citenamefont{Feldman and Nelson}(2005)}]{feldman}
\bibinfo{author}{\bibfnamefont{B.}~\bibnamefont{Feldman}} \bibnamefont{and}
  \bibinfo{author}{\bibfnamefont{A.}~\bibnamefont{Nelson}}
  (\bibinfo{year}{2005}).

\bibitem[{\citenamefont{Takahashi and Tanimoto}(2005)}]{Takahashi:2005kw}
\bibinfo{author}{\bibfnamefont{R.}~\bibnamefont{Takahashi}} \bibnamefont{and}
  \bibinfo{author}{\bibfnamefont{M.}~\bibnamefont{Tanimoto}}
  (\bibinfo{year}{2005}), \eprint{hep-ph/0507142}.

\end{thebibliography}
\bibliographystyle{apsrev}
\end{document}